\documentclass[twocolumn,aps,prb,superscriptaddress,showpacs]{revtex4}
\usepackage{graphicx,amscd,amsmath,amssymb}
\begin{document}
\title{Low Energy Coherent Transport in Metallic Carbon Nanotube Junctions}
\author{A.A. Maarouf}

\affiliation{IBM T.J. Watson Research Center \\
    Yorktown Heights, NY 10598 \\
}
\affiliation{
Egypt Nanotechnology Research Center \\
Smart Village, Building 121, Giza, Egypt 12577, 
}

\author{E.J. Mele}
\affiliation{
Department of Physics and Astronomy, University of Pennsylvania \\ 
Philadelphia, PA 19104}

\date{\today}
 
\begin{abstract}
We study the low-energy electronic properties of a junction made of two crossed metallic carbon nanotubes of general chiralities. We derive a tight binding tunneling matrix element that couples low-energy states on the two tubes, which allows us to calculate the contact conductance of the junction. We find that the intrinsic asymmetries of the junction cause the forward and backward hopping probabilities from one tube to another to be different. This defines a zero-field Hall conductance for the junction, which we find to scale inversely with the junction contact conductance. Through a systematic study of the dependence of the junction conductance on different junction parameters, we find that the crossing angle is the dominant factor which determines the magnitude of the conductance.

\end{abstract}


\maketitle

\section{Introduction}

Since their discovery almost two decades ago\cite{iijima1}, carbon nanotubes (CNTs) have been the subject of intense experimental and theoretical research. CNTs come in different geometries: single wall tubes, multiwall tubes, and single wall tube bundles (ropes). Different CNTs are characterized by their chirality, which is a measure of how the tube lattice is oriented with respect to the tube axis.

In order to understand the bulk behavior of CNT systems, the electronic properties of different CNT geometries have been studied\cite{revmodphys}, with some focus on the multiwall tubes \cite{ando1,sglouie,Mayou, ando2, chang2,Uryu,rubio}. These studies addressed the issue of intertube transport and its dependence on the chiralities of the tubes and the geometrical details. Most recently, films of CNTs have been used as a transparent conducting electrode\cite{TCE}. Such films are made of networks of CNT ropes. In a rope, the constituent tubes have a distribution of chiralities. The resistance of these films is dominated by the tube-tube contacts(junctions). At such junctions, tubes with different chiralities intersect with different crossing angles. Therefore, it is desirable to study how these two factors (tube chirality and crossing angle) affect the transport properties of these junctions. 

Junctions of crossed CNTs have been studied experimentally\cite{fuhrer2,device1}.  In their work, Fuhrer et al.\cite{fuhrer2} explored the different electrical properties of different combinations of crossed metallic and semiconducting tubes. Metal-metal junctions showed a contact conductance of the order of $0.02 (4e^2/h)$, despite the small junction area. This was attributed to the elastic deformation of the tube arising from the interaction with the substrate, resulting in an intertube distance smaller than the expected $3.4 {\mbox \AA}$. This increases the coupling at the contact region, and provides a natural explanation of the relatively high measured conductance. Crossed Nanotube junctions have been theoretically studied for high symmetry cases only\cite{ando3, buldum, buia}, where it was found that maximal contact conductance occurs when the tube lattices are commensurate. Despite such  effort, the general problem of two crossed CNTs has not yet been considered.  Of interest is the effect of tube chiralities, the crossing angle as well as the relative orientation of the tube lattices on the transport properties of these junctions.

In this paper we present a general study of that problem in a tight binding framework. We derive a tunneling matrix element that couples low energy propagating states on each tube. This matrix element depends on the chiral angles, the crossing angle, and the Fermi level of the junction. This allows us to systematically study the effects of the different junction parameters on the junction conductances. We find that due to the intrinsic asymmetries of the junction, forward and backward tunneling between one tube and the other are generally unequal. Therefore, passing a current in one tube leads to the development of a non-zero voltage across the other one, thereby defining a zero-field Hall-like conductance for the junction. Furthermore, we find that this zero-field Hall conductance relates simply to the contact conductance of the junction. Our study also shows that the electronic properties of the junction sensitively-depend on the degree of matching between the tube lattices. For a given junction, this matching is controlled by the crossing angle, which results in an intertube conductance that varies  by an order of magnitude for different angles. 

The paper is organized as follows. In section II we introduce our tunneling model and derive the tunneling matrix element coupling low energy tube states. In section III we derive a formula for the junction conductance in terms of the microscopic conductances of the junction in a Landauer-B{\"u}ttiker framework, and present some numerical results showing the dependence of the junction conductances on various junction parameters. The paper is concluded  in section IV.
 
\section{Tunneling Model}

\begin{figure}
   \includegraphics[width=1.5in]{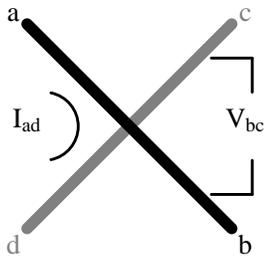}
   \caption
{ 
Two crossed metallic tubes with different chiralities, and a crossing angle $\beta$.
}

   \label{setup0}
\end{figure}

\begin{figure}
   \includegraphics[width=2.5in]{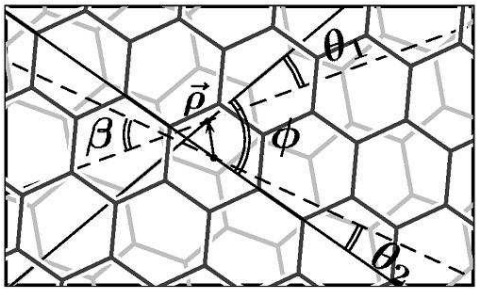}
   \caption
{Details of the contact region. The dotted lines define the axes of the crossed tubes, with chiral angles $\theta_1$ and $\theta_2$. $\beta$ is the crossing angle. Tubes face each other from the outside. The mismatch between the two lattices is parametrized by an angle $\phi=\theta_1 + \theta_2 + \beta$. The 2-D vector ${\boldmath \rho}$ defines the relative orientation of the origins of the two lattices(which are taken to be hexagon centers).
}
\label{setup}
\end{figure}

In this section we describe the  low-energy coupling between two crossed tubes (Fig. \ref{setup0}) within  a tight binding formalism.  Nanotubes are assumed to be long and free of defects. The overlap region is shown in Fig. \ref{setup}, where $\beta$ is the crossing angle, ${\boldmath \rho}$ is a vector defining the relative displacement of the origins(defined as hexagon centers) of the two lattices, and $\theta_1$ and $\theta_2$ are the chiral angles. Matching between the two lattices can by quantified by a {\it registry angle} $\phi$  defined by
\begin{equation}
\phi = \theta_1 + \theta_2 + \beta
\end{equation}

The uncoupled tubes are described by a nearest-neighbor tight-binding Hamiltonian
\begin{equation}
{\cal H}_0 = -\sum_{a=1,2}\sum_{\big < ij \big >} t_{\pi} c^{\dagger}_{ai} c_{aj} 
\end{equation}
where the index $a$ labels the CNTs and $\big <ij\big>$ is a sum over nearest-neighbor atoms on each CNT. The eigenstates of ${\cal H}_0$ are plane waves localized on each CNT. The interaction Hamiltonian ${\cal H}_T$ is built in such a way that an electron can hop from any atom on one CNT to any atom on the other
\begin{equation}
 {\cal H}_T=\sum_{ij} t_{ij} c^{\dagger}_{1i} c_{2j},
\end{equation}
where $t_{ij}$ depends on the positions and relative orientation of the $\pi$ orbitals on atoms $i$ and $j$, and varies exponentially with the distance between the two hopping sites 
\begin{equation}
t_{ij} = t_0 e^{-d_{ij}/ a_0},
\end{equation}
where $t_0$ is a free parameter to be determined,  and $a_0=0.529 \AA$ is the range of the $\pi$ orbitals. The inter-site distance $d_{ij}$ is given by
\begin{eqnarray}
d_{ij}^2  & = & (b + 2R-R \cos\frac{y_1}{R} - R \cos\frac{y_2}{R})^2  \nonumber \\ & + &  (z_2 \sin\beta+R \sin\frac{y_1}{R} - R \sin\frac{y_2}{R} \cos\beta)^2  \nonumber \\ & + & (z_2 \cos\beta-z_1 - R \sin\frac{y_2}{R} \sin\beta)^2,
\end{eqnarray}
where $z_a$($y_a$) is the distance along the length (waist) of CNT $a$, $b$ is the intertube separation at the point of closest contact. The CNTs are assumed to be of approximately equal radii, which is denoted by $R$. We will also assume that hopping between the CNTs is dominated by $y_a,z_a \ll b,R$. The total Hamiltonian of the system 
\begin{equation}
{\cal H} = {\cal H}_0+{\cal H}_T,
\end{equation}
is then expressed in a plane wave basis by the transformation 
\begin{equation}
c_{ai}= \frac{1}{\sqrt{N}} \sum_{{\bf k}_a}e^{i {\bf k}_a \cdot {\bf r}_{ai}} c_{a\eta(i){\bf k}_a},
\end{equation}
 where ${\bf r}_{ai}$ is the position vector of site $i$ on tube $a$, $\eta$ specifies the $A$ or $B$ sub-lattice, $N$ is the number of graphene unit cells in the tube, and ${\bf k}_a = (k_{ay},k_{ax})$. In this basis 
\begin{equation}
{\cal H}_0 = -t_\pi \sum_{a=1,2}\sum_{{\bf k}_a} \gamma_{{\bf k}_a} c_{aA {\bf k}_a}^{\dagger} c_{aB {\bf k}_a} + H.c.,
\end{equation}
where 
\begin{equation}
\gamma_{{\bf k}_a} = \sum_{j=1}^3 e^{i {\bf k}_a \cdot {\bf d}_{aj}},
\end{equation}
and ${\bf d}_{aj}$ are the three nearest-neighbor vectors connecting the two sub-lattices of tube $a$.

\begin{figure}
\includegraphics[width=3.2in]{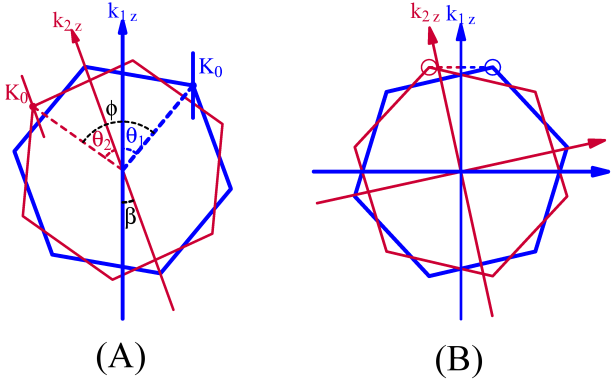}
\caption
{
Brillouin zones of the lattices of the two crossed tubes. {\bf A}. The zones are rotated with respect to each other due to the difference in chiral angles of the tubes, as well as the crossing angle $\beta$. The axial momentum directions of the two tubes are shown. The rotation of one tube lattice with respect to the other is quantified by the angle $\phi$. {\bf B} The zones of the two lattices at one of the angles $\beta^-$ where the tunneling is dominated by pairs of Fermi points with zero axial momentum mismatch is zero and finite azimuthal momentum one.
}
\label{BZ}
\end{figure}
The Brillouin zones of the graphene sheets forming the two crossed tubes are shown in Fig. \ref{BZ}A. The wrapping of a graphene sheet corresponds to slicing the 2-D zone in a direction parallel to the tube axis. This yields a series of 1-D bands, and in the case of a metallic tube, two bands cross at the Fermi energy.    At low energy, one focuses on the neighborhood of the Fermi points(corners of the Brillouin zone), where ${\bf k} = \alpha {\bf K}_p + {\bf q}$, ${\bf q}=(0,q)$, $\alpha=\pm 1$, and $p=-1,0,1$ denote the three equivalent Fermi points. In this limit, the tube Hamiltonian is  diagonalized by the transformation 
\begin{equation}
U=e^{-i (1/2) \alpha \theta \sigma^z} e^{-i (\pi/4) \alpha \sigma^y},
\end{equation}
where $\sigma^{y,z}$ are the Pauli matrices, and we get the two eigenmodes $\psi_{Rq}$ (right movers, R) and $\psi_{Lq}$ (left movers, L). In this R/L basis ${\cal H}_0$ becomes
\begin{equation}
{\cal H}_0=\sum_{a=1,2}\sum_{q_a} v_F q_a (\psi^{\dagger}_{aR{\boldmath q}_a} \psi_{aR {\boldmath q}_a} -  \psi^{\dagger}_{aL{\boldmath q}_a} \psi_{aL{\boldmath q}_a}),
\end{equation}
with $v_F= 3 t_\pi d/2$ and $d$ is the nearest neighbor distance. The tunneling Hamiltonian ${\cal H}_T$ can be expressed in the propagating states basis by following a similar procedure. Introducing the plane wave basis in ${\cal H}_T$ gives
\begin{eqnarray}
{\cal H}_T & = & \sum_{\eta_1 \eta_2} \sum_{{\bf G}_1 {\bf G}_2 {\bf k}_1 {\bf k}_2}
e^{i{\bf G}_1\cdot (\rho_1 + \eta_1 \tau_1) - {\bf
G}_2\cdot(\rho_2 + \eta_2 \tau_2)}  \nonumber \\ & \times & t_{{\bf k}_1+{\bf G}_1, {\bf k}_2 + {\bf G}_2} c_{\eta_1{\bf k}_1}^\dagger  c_{\eta_2{\bf
k}_2} + H.c.,
\label{G}
\end{eqnarray}
where ${\bf k}_i = \alpha_i {\bf K_0}_i + {\bf q}_i$, ${\boldmath \rho}_i$ are 2-D vectors defining the origins of each graphene lattice from the point of closest contact, $\eta_i$ are $\pm 1$, $\tau_i$ are vectors joining the origin of each lattice to the nearest $A$ sub-lattice site, and $t_{{\bf k}_{1},{\bf k}_{2}}$ is determined from the 2-D Fourier transform of the spatial hopping element $t_{ij}$, and is given by
\begin{equation}
t_{{\bf k}_{1},{\bf k}_{2}}  = t_J \frac{e^{-\frac{a_{0}}{2}(b+2R)f_1}  e^{a_{0}R f_2}}{L |\sin\beta|} ,
\label{tkk}
\end{equation}
where
\begin{equation}
f_1 = \frac{k_{1z}^{2} + k_{2z}^{2} - 2k_{1z}k_{2z}\cos\beta}{\sin^{2}\beta},
\label{f1}
\end{equation}
\begin{eqnarray}
f_2 & = & \frac{k_{1y}(k_{2z} - k_{1z}\cos\beta)+k_{2y}(k_{1z}-k_{2z}\cos\beta)}{\sin\beta} \nonumber \\ & + & \frac{1}{2}((k_{1z}^{2} + k_{2z}^{2}) - (k_{1y}^{2} + k_{2y}^{2})),
\label{f2}
\end{eqnarray}
\begin{equation}
t_J = \frac{2 \pi a_{0}^{2} b}{A} t_0 e^{-\frac{b}{a_{0}}}.
\end{equation}
$A$ is the area of a graphene unit cell, $L$ is the length of the tubes, and $t_J$ is an overall magnitude of the $k$-space tunneling matrix element. This matrix element reproduces the result obtained for the case of parallel tubes(in the limit $\beta \rightarrow 0$)\cite{maarouf}. Taking the limits $\beta \rightarrow 0$ and $R \rightarrow \infty$ gives us the hopping  matrix element between two graphene sheets. This result can be used to calculate the low energy spectrum of two graphene sheets with Bernal stacking, as well as the transverse bandwidth of graphite\cite{maarouf}. Fitting $t_J$ to the former, and for $b=3.4\AA$, we get $t_J= 0.35$ eV . 

We then express the tunneling Hamiltonian in the R/L moving basis. To do this, we notice from Eqs. (\ref{tkk}-\ref{f2}) that $t_{{\bf k}_{1},{\bf k}_{2}}$ depends on the magnitudes of the momenta of the initial and final states, and decays rapidly as $|{\bf k}|$ increases. Therefore the sum over reciprocal lattice vectors in Eq. \ref{G} is dominated by the first star. ${\cal H}_T$ thus becomes
\begin{equation}
{\cal H}_T = \sum_{\alpha_1 \alpha_2 \sigma_1 \sigma_2 q_1 q_2}T(\alpha_1 \sigma_1 {q}_1 | \alpha_2 \sigma_2 {q}_2) \psi^{\dagger}_{1 \alpha_1 \sigma_1 {q}_1} \psi_{2 \alpha_2 \sigma_2 {q}_2} + H.c.
\end{equation}
where $\sigma=R,L$, and $T(\alpha_1 \sigma_1 {q}_1 | \alpha_2 \sigma_2 {q}_2)$ is given by
\begin{eqnarray}
T(\alpha_1\sigma_1{\bf q}_1 | \alpha_2\sigma_2 {\bf q}_2)
& = & \sum_{\ell_1 \ell_2=-1}^1 e^{ i \alpha_1 {\bf K}_{1 \ell_1} \cdot
\rho_1 - i \alpha_2 {\bf K}_{2 \ell_2} \cdot \rho_2 } \nonumber \\ & \times & t_{\alpha_1 {\bf K}_{1 \ell_1}+ {\bf q}_1, \alpha_2{\bf
K}_{2 \ell_2} + {\bf q}_2} M_{\sigma_1\sigma_2},
\label{bigT}
\end{eqnarray}
where
\begin{equation}
M =
 {1\over 2} \left[ \begin{array}{*2c}
          f_{\alpha_{1}}^{\ell_1}   f_{\alpha_{2}}^{\ell_2 \ast}
      &  f_{\alpha_{1}}^{\ell_1}   f_{-\alpha_{2}}^{\ell_2 \ast} \\
 f_{-\alpha_{1}}^{\ell_1}   f_{\alpha_{2}}^{\ell_2 \ast}
      &  f_{-\alpha_{1}}^{\ell_1}   f_{-\alpha_{2}}^{\ell_2 \ast}
\end{array} 
\right],
\label{geo1}
\end{equation}
\begin{equation}
f_{\alpha}^{\ell} = e^{i \zeta_\ell} + \alpha e^{- i \zeta_\ell},
\label{geo2}
\end{equation}
and
\begin{equation}
\zeta_\ell = {2\pi\ell \over 3} - {\theta \over 2}.
\label{zeta}
\end{equation}

The form of $t_{{\bf k}_{1},{\bf k}_{2}}$ (Eq. \ref{tkk}-\ref{f2}) shows that the coupling between propagating modes on the tubes depends on the geometry of the junction. In addition, the Fermi energy of the system has an effect on the coupling, as the momentum difference between propagating states on the two tubes changes as the Fermi energy is changed.

It should be noted here that our developed theory is a low energy one, and it applies where the linearization of the tube band structure is valid. An upper limit of the energy range (around the Dirac point) where our model is applicable is determined by the diameters of the tubes forming the junction. As an example, for tubes with diameters of $\sim 1.4$ nm, the higher sub-bands enter the picture at $\sim 0.8$ eV above and below the Dirac point\cite{saitokataura}. This energy scale decreases with increasing tube diameter ($\sim 0.3$ eV for a diameter of $\sim 4$ nm). The low energy limit for the applicability of our model is imposed by the curvature gaps of small diameter tubes ($\sim$ 0.5 nm),where the $\pi$-electron tight binding description fails to describe the band structure correctly close to the Dirac points\cite{magna, chico}. These gaps are in the range of $50$ meV for a diameter of about $0.7$ nm\cite{curvgaps1,curvgaps2}.

With Eqs. (\ref{tkk}-\ref{zeta}), we have a model that describes  electronic coupling between two crossed metallic tubes. The virtue of our results is that they can be used to study the intertube conductances as a function of tube chiralities, crossing angle, Fermi energy, as well as the relative orientation of the tube lattices. Such a study is presented in the next section.

\section{Junction Conductances}

In this section, we derive a few formulas for the junction conductances. A thorough treatment of the fundamentals of mesoscopic transport is given by Datta\cite{transport_book}.

One can abstractly view the two crossed tubes as a four-terminal device (Fig.\ref{setup0}). Coherent transport in such devices has been studied before \cite{Buttiker}. In a four-terminal system with time reversal invariance, three different resistance measurements can be made. In Fig. \ref{setup0}, one can imagine passing a current between terminals $a$ and $d$, and measuring the voltage across terminals $c$ and $b$, thus defining a resistance $R_{ad,cb}$. Similarly, another two resistances, $R_{ac,db}$ and $R_{ab,dc}$, can be measured.  The three resistances are subject to the simple constraint $R_{ad,cb} + R_{ac,db} + R_{ab,dc} = 0$ \cite{Buttiker}. The forms of these resistances depend on the relations between different microscopic conductances of the system under study. According to the geometry of the system, one of these three resistances can be thought of as a zero-field Hall resistance($R_{ab,dc}$). Four-terminal semiconductor hetero-structures have been previously studied \cite{Baranger, Chang}. The Hall resistance measured in such systems is due to an asymmetry induced by impurities. The CNT junction we study here is fundamentally different from those since the asymmetry causing the zero-field Hall resistance is an intrinsic property of the junction. We will now derive analytic results for the junction resistances in the Landauer-B{\"u}ttiker formalism, and using our results from the last section, we will study the dependence of these resistances on different junction parameters.

The transmission matrix  between  propagating states of different tubes can be calculated using Eqs. (\ref{tkk}-\ref{zeta})  through Fermi's Golden rule. This allows us to calculate different intertube conductances. For linear response, these conductances are:
\begin{equation}
G_{\sigma \sigma'} =  4\pi^2 \rho^2_{F} \sum_{\alpha \alpha'}  \big | \big <\sigma \alpha |{\cal H}_{T}| \alpha' \sigma'\big > \big |^{2} G_0,
\end{equation}
where $\sigma,\sigma'=R,L$, $\rho_F$ is the density of states at the Fermi energy, and $G_0=2e^2/h$. Because of time reversal symmetry, $G_{LL}=G_{RR}$ and $G_{LR}=G_{RL}$, and therefore there are only two different intertube conductances, denoted by $G_f$ and $G_b$, respectively. In addition, in the absence of scattering, we have the intrinsic conductances of the tubes, $G_i=2G_0, i=1,2$. As we will see, if the distance between the two tubes is not very different from that in a bundle, coupling between the tubes will be weak,  or equivalently, $G_f,G_b \ll G_1,G_2$. 

As pointed out, three conductance measurements can be done for the crossed tubes device.  $G_c^I=R_{ac,bd}^{-1}$ is the first contact conductance of the junction, and is determined  by passing a current between terminals $a$ and $c$ and measuring the voltage across the $b$ and $d$(see Fig. \ref{setup0}). $G_c^{II}=R_{ad,bc}^{-1}$ is the other contact conductance of the junction, and is measured the same way as $G_c^I$ but with $c$ and $d$ interchanged. The third conductance of the system, $G_H=R_{ab,cd}^{-1}$, is a Hall-like conductance, where a current is passed in one tube and a voltage is measured across the other one. We use the B{\"{u}}ttiker formula to obtain the following expressions for the three conductances:
\begin{equation}
\frac{1}{G_c^{I,II}}  =  \frac{1}{2(G_f + G_b)} - \frac{1}{2(2G_0-G_{b,f}) }
\end{equation}
and
\begin{equation}
\frac{1}{G_H} =  \frac{(G_f - G_b)}{2(2G_0-G_f)(2G_0-G_b)}
\end{equation}
The three conductances satisfy a simple constraint \cite{Buttiker}, ${G_c^{I}}^{-1}-{G_c^{II}}^{-1}=G_H^{-1}$. Since we expect $G_{f,b} \ll G_0$, we can approximate the above formulas by
\begin{equation}
\frac{1}{G_c^{I,II}} = {1 \over 2}{ 1 \over G_f + G_b} - {1 \over 4 G_0} - {G_{b,f} \over 8 G_0^2},
\end{equation}
and
\begin{equation}
\frac{1}{G_H} = \frac{(G_f - G_b)}{8 G_0^2} 
\end{equation}
where now the two contact conductances are equal to leading order in $G_f/G_0$ and $G_b/G_0$, and we will denote them by $G_c$. The product $G_c G_H$ then becomes(to leading order in $G_{f,b}/G_0$)
\begin{equation}
 G_c G_H = \frac{1}{\eta} \Big(\frac{8e^2}{h}\Big)^2
\label{mainres}
\end{equation}
where $\eta = (G_f - G_b)/ (G_f + G_b)$. Since $G_f$ and $G_b$ are generally of the same order, $\eta$ will be of order unity. Being a ratio between the difference and sum of the intertube conductances, the proportionality parameter $\eta$ is independent of the tunneling strength $t_0$. Therefore, in such a system the Hall conductance is of order of the reciprocal of the contact conductance. This result is not restricted to our present system, but rather, it is a general result for any four terminal  device which possesses time reversal symmetry, with two of its conductances much smaller than the other two.

The zero-field Hall voltage that develops across a tube upon passing current in the second one is an intrinsic property of the tube junction. It is a manifestation of the difference between forward and backward transmission probabilities. Because of the Gaussian form of the matrix element(\ref{tkk}), it turns out that (for most chiralities and crossing angles) tunneling will be dominated by one set of Fermi points, say, ${\bf K}_{10}$(of first tube) and ${\bf K}_{20}$(of second tube). In that case, the tunneling matrix elements are:
\begin{eqnarray}
T(+Rq_1 | + R q_2) & = & 2 e^{i{\bf K}_{10} \cdot {\boldmath \rho_1 }- i {\bf K}_{20} \cdot {\boldmath \rho_2} }t_{+ {\bf K}_{10}+{\bf q}_1 , +{\bf K}_{20} + {\bf q}_2} \nonumber \\ & \times & \cos{ \theta_1 \over 2}  \cos{ \theta_2 \over 2},
\label{TRR}
\end{eqnarray}
and
\begin{eqnarray}
T(+Rq_1 | + L q_2) & = & -2 i e^{i{\bf K}_{10} \cdot {\boldmath \rho_1 }- i {\bf K}_{20} \cdot {\boldmath \rho_2} } t_{+ {\bf K}_{10}+{\bf q}_1 , +{\bf K}_{20} - {\bf q}_2} \nonumber \\ & \times & \cos{ \theta_1 \over 2}  \sin{ \theta_2 \over 2}.
\label{TRL}
\end{eqnarray}

\begin{figure}
\includegraphics[width=2.5in]{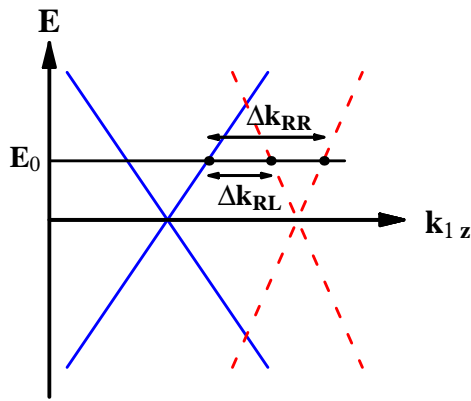}
\caption
{
Low energy band structure of two metallic tubes of different chiralities in the neighborhood of one set of K-points. The $x$-axis is along the first tube. The first tube bands (solid) and the second tube ones (dashed) are shown as a function of the axial momentum of the {\it first} tube, $k_{1x}$, so as to be able to compare the axial momentum mismatch of states at the same energy $E$. This results in the bands of the second tube being weighed by $\cos \beta$. At an energy $E_0$, the momentum mismatch for scattering between right moving states, $\Delta k_{RR}$, is different from that between right and left  moving ones, $\Delta k_{RL}$.
}
\label{bands}
\end{figure}
The discrepancy between  $T(+ R q_1 | + R q_2)$ and   $T(+ R q_1 | + L q_2)$ arises from the momentum mismatch between the initial and final tube states. To understand this more, one must have a closer look at the tube bands near the Fermi level. In Fig. \ref{bands} we show the lowest lying bands of a metallic tube(solid), and the same-energy states of another metallic tube of a different chirality(dotted). The horizontal axis is the momentum along the axis of the first tube, $k_{1z}$. The second tube states(dashed) are plotted as a function of $k_{1z} $ as well. Tubes with different chiralities have different band-crossing points. The position of the Fermi level defines the axial momenta of the left and right moving states on each tube. The tunneling matrix element depends on terms of the form:
\begin{equation}
\Delta k = k_{1z} - k_{2z} \cos\beta,
\end{equation}
which is the difference between the axial momentum of one tube and the projection of the axial momentum of  the second tube on the first one. At an energy $E_0$,  the quantity $\Delta k$  is given by: 
\begin{equation}
\Delta k_{RL,RR} = {E_0 \over v_F} (1 \pm \cos \beta) + ({\bf K}_{10})_{z} - ({\bf K}_{20})_{z} \cos \beta
\label{mom}
\end{equation}
and therefore $t_{{\bf k}_1 , {\bf k}_2}$ will be different for forward and backward scattering.

It is worthwhile comparing the geometrical effects in the crossed tubes case to our previous work of two parallel tubes\cite{maarouf}.  For parallel tubes, the tunneling matrix element imposes  axial-momentum conservation due to the effective translational symmetry of the system, and {\it approximate} momentum conservation in the azimuthal direction through its Gaussian dependence on the difference between the final and initial states azimuthal components. For the case of crossed tubes, we have {\it approximate} 2-D momentum conservation; maximum coupling occurs when the 2-D momentum mismatch is minimum, which corresponds to highest matching between the tube lattices. We can summarize the effect of the crossing angle on the conductance in three cases. First, a general case where there is a finite 2-D momentum mismatch between the tube states(Fig. \ref{BZ}A). The greater this mismatch, the more suppressed is the inter-tube conductance. The second case is when the crossing angle is such that the Brillouin zones of the two lattices overlap, i.e. when $\phi=n\frac{\pi}{3}, n=0,1,2$, which will occur at crossing angles (defining $\beta$ in the clockwise direction) 
\begin{equation}
\beta_n^+=n\frac{\pi}{3}+(\theta_1 + \theta_2).
\label{star1}
\end{equation}
At such angles, the 2-D momentum mismatch between the initial and final tube states is minimal. The third case is when the momentum mismatch is zero for one component only(Fig. \ref{BZ}B). This occurs at crossing angles given by
\begin{equation}
\beta_n^-=n\frac{\pi}{3}+(\theta_1 - \theta_2).
\label{star2}
\end{equation}
In general, conductance peaks occurring at these angles will be of lesser magnitude than those occurring at $\beta_n^+$. Therefore, we expect to see three peaks in the angular dependence of the intertube conductance corresponding to the three angles $\beta_n^+$ where the tube lattice commensuration is maximal, with some fine structure arising at the angles $\beta_n^-$.

 Another effect of the crossing angle is that it controls the effective contact area between the two lattices, which is minimum at $\beta=\pi/2$. Therefore, one expects conductance peaks occurring close to a crossing angle of $\pi/2$ to be of a smaller magnitude than those occurring farther away from $\pi/2$.

To illustrate  the dependence of the junction conductances on the different junction parameters, we study the roles of the chiral angles, crossing angle, relative orientation of the tube lattices, and the Fermi level of the junction. We perform this study for many  junctions, of which we choose to show only three. The first ({\bf I}) is a (17,2)-(10,10) junction, the second ({\bf II}) is a (17,2)-(15,6), and the third ({\bf II}) is a (13,7)-(13,7) one. The first two represent the general case of two tubes with different chirality, while junction {\bf III} represents a junction which may be prepared experimentally by AFM manipulation\cite{manipulation}. We take the intertube separation $b$ to be similar to  that of tubes in a bundle, $b=3.4  {\mbox \AA}$. The effect of the substrate would be to decrease the intertube separation. This can be easily incorporated into our model by using an effective separation $b_{eff} < 3.4  {\mbox \AA}$. For the relative orientation of the CNT lattices, we study two limiting cases. The first is when the point of closest contact between the CNTs is two hexagon centers(HH-orientation), and the second is when that point is between an atom and a hexagon center (AH-orientation).

\begin{figure}
\includegraphics[width=3.2in]{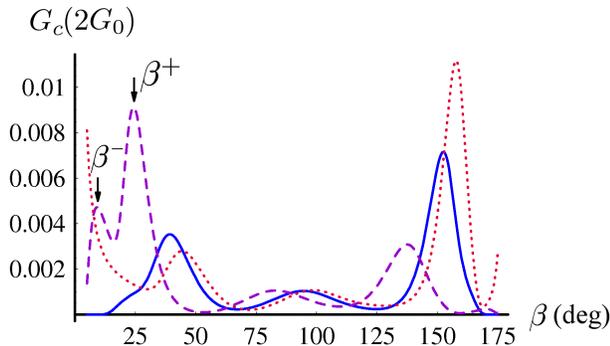}
\caption
{
Dependence of $G_c$ on the crossing angle $\beta$  for three junctions; {\bf I}  (17,2)-(10,10), {\bf II} (17,2)-(15,6)(dashed), and {\bf III} (13,7)-(13,7)(dotted), for $E_0=0$. Different peaks in each plot mark the angles where there is high registry between the two tube lattices. The two arrows mark the angles $\beta^-$ and $\beta^+$ for junction {\bf II}.  
}
\label{whatever}
\end{figure}

\begin{figure}
\includegraphics[width=3.0in]{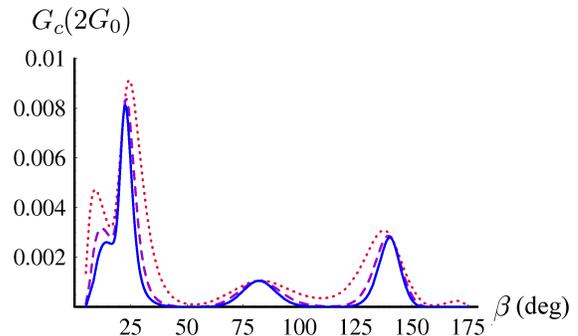}
\caption
{
Dependence of $G_c$ on the crossing angle $\beta$ for 3 junctions. The radii of the tubes in the 3 junctions are R(solid), 2R(dashed), and 3R(dotted), and $R \sim 1.5$ nm. The CNTs are assumed to have an HH-orientation contact, and the Fermi energy $E_0=0$.
}
\label{radiusdependence}
\end{figure}

We begin by discussing the dependence of the contact conductance, $G_c$, on different parameters. Figure \ref{whatever} shows the angular dependence for $G_c$ for the three junctions. The Fermi energy, $E_0=0$, is at the ${\bf K}$ point, and the CNT lattices are oriented with an HH orientation. For junction {\bf I} (solid plot), we see that there are three main peaks in the conductance, which can be related to maxima in the transmission probabilities between the two tubes at these crossing angles. As argued before, and since the tubes are of different chirality, there exist certain crossing angles at which lattice commensuration occurs, and in such cases the transmission probabilities will be largest, leading to a conductance maximum. According to Eq. \ref{star1}, the crossing angles where we should see such peaks are $\beta^+ =36^\circ,96^\circ,$ and $156^\circ$ for junction {\bf I}, which is what we see in Fig. \ref{whatever}. On the other hand, crossing angles which maximize the lattice mismatch will result in a contact conductance which is roughly  an order of magnitude lower. Junction {\bf II} shows a qualitatively-similar behavior(Fig \ref{whatever} dashed plot). For this junction, lattice matching occurs at $\beta^+=22^\circ, 82^\circ$, and $142^\circ$. For junction {\bf III}, $\beta^+=40^\circ,100^\circ,$ and $160^\circ$. We also see that for this junction the conductance increases as $\beta \rightarrow 0^\circ, 180^\circ$ as the system becomes periodic and the tunneling matrix element takes a $\delta$-function form \cite{maarouf}. 

The discrepancy in the magnitudes of the three conductance peaks (for each junction) can be understood in terms of the contact area of the junction. The transmission probability, and hence the conductance, increases with the area of overlap between the two lattices. This area is minimum at perpendicular crossing, which makes the conductance peaks closer to $\beta=0^{\circ},180^{\circ}$ relatively larger than those near $\beta = 90^{\circ}$.

The slight deviation of the conductance peaks from the lattice-commensuration angles that we see in Fig. \ref{whatever} is due to the finiteness of the tube radii. For a radius of $\sim 1.5$ nm, the curvature causes a deviation of a couple of degrees. This deviation decreases with increasing tube radii. In addition, this deviation is smallest at perpendicular crossing, as the curvature effects are minimal. We show that in Fig. \ref{radiusdependence}, for junction {\bf II}(dotted plot) and two other junctions of the same chirality but with bigger radii; a (34,4)-(30,12) junction (radii $\sim 3$ nm, dashed), and a (51,6)-(45,18) junction (radii $\sim 4.5$ nm, solid). As the radii increase, deviation from the angles $\beta^+$ becomes negligibly small. We also see that the peaks occurring at angles $\beta^-$ get increasingly small as the radii of the tubes increase. These are the angles at which there is a finite mismatch in only one of the 2D-momentum components, which causes these peaks to be suppressed as the tube radii are increased.
\begin{figure}
\includegraphics[width=3.0in]{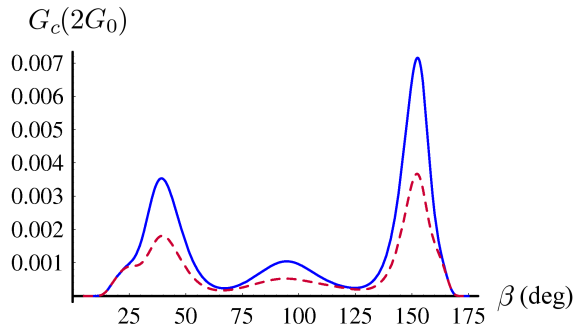}
\caption
{
Dependence of $G_c$ on the crossing angle $\beta$  for junction {\bf I}(17,2)-(10,10) with HH-(solid) and AH-(dashed) orientations at $E_0=0$eV.
}
\label{relativeorient}
\end{figure}

The dependence of the conductance on the relative orientation of the tube lattices is investigated in Fig. \ref{relativeorient}, where we show the angular dependence of the conductance of junction {\bf I} for two orientations; HH, and AH. A first observation is that the relative orientation has a negligible effect on the positions of the conductance peaks(as should be the case since these positions are a function of the Fermi-points mismatch). In addition, it is clear that the HH orientation gives a higher conductance, specially at the peaks where the lattice commensuration occurs. This is expected as in the HH orientation more lattice sites are matched, compared to the AH one. 
\begin{figure}
\includegraphics[width=3.0in]{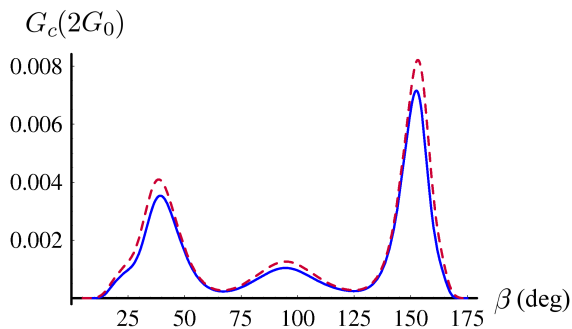}
\caption
{
Dependence of $G_c$ of junction {\bf I}(17,2)-(10,10) on the crossing angle $\beta$  for different energies: $E_0=0$eV(solid), and $E_0=-0.2 eV$(dashed).
}
\label{condgate}
\end{figure}

Whereas varying the crossing angle of a certain junction can be  a challenging experimental task, changing the Fermi level should be easily achievable through electrostatic doping. Therefore, the Fermi level dependence of the contact conductance is of interest. In Fig. \ref{condgate}, we show the angular dependence of the contact conductance for junction {\bf I} for different energies. We find that whereas changing the Fermi level has a negligible effect on the position of the conductance peaks, it does affect the magnitude of the conductance. This can be understood in band structure terms. Tuning the Fermi level away from the ${\bf K}$ points changes the momentum mismatch between the initial and final states involved in the tunneling(Eq. \ref{mom}), thereby changing the transmission probability and the conductance. 
\begin{figure}
\includegraphics[width=2.5in]{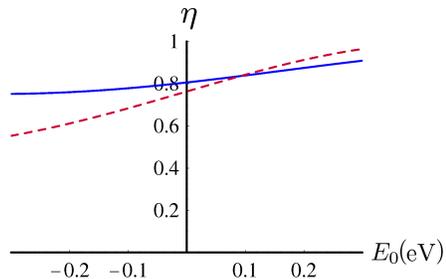}
\caption
{
Dependence of $\eta$ on the Fermi energy for junction {\bf I}(17,2)-(10,10)(solid) and junction {\bf III}(13,7)-(13,7) (dashed) at $\beta=40^\circ$ with HH orientation.
}
\label{eta17}
\end{figure}
Another quantity which can be probed experimentally is $\eta$. The virtue of such measurement is that, being  readily predictable from our theory, it provides a direct test for it. As mentioned before, $\eta$ is independent of the intertube separation. Our model predicts $\eta$ to be of order unity. The variation of $\eta$ with the Fermi level is also achievable experimentally. In Fig. \ref{eta17}, we show the variation of $\eta$ with the Fermi level for junctions {\bf I}(solid) and {\bf III}(dashed), for $\beta =40^{\circ}$. As we see, $\eta$ is of order unity. We find this to be the case for most junctions. A sign change of $\eta$ reflects the fact that the Fermi level affects the forward and backward conductances differently. The unique point in parameter space that has $\eta=0$ is a point of high symmetry in the sense that the many asymmetries of the junction counter-act to give equal forward and backward transmission probabilities.

Although electron-electron interactions are not included in our treatment, Eq. (\ref{mainres}) still holds. Such interactions tend to renormalize the tunneling density of states\cite{kane2}; $\rho_F^{int} = \rho_F (\frac{T}{T_F})^\alpha$, and therefore $\eta$ would not change if such interactions were included. The estimates  $\alpha \sim 0.6$ and $T_F \sim 1$eV give  $\rho_F^{int} \sim 0.1 \rho_F$ at room temperature. Another effect that we have ignored is the small deformation occurring in the tubes at the contact region, as predicted by molecular dynamics simulations \cite{hertel, rochefort}. Such deformations slightly increase the contact area between the two tubes, and may also cause the on-tube conductance to be slightly lower than the assumed $2G_0$ value due to possible back-scattering. These two effects are expected to be of more relevance for large diameter tubes, where faceting at the contact region is more pronounced. Detailed treatment of the deformation effects requires a microscopic approach using a real space Hamiltonian\cite{neto}.

\section{Conclusion}

In this paper, we study the low-energy electronic properties of a junction formed by two crossed metallic CNTs in a tight-binding framework. We derive a tunneling matrix element that couples the low-energy electronic states on the two tubes. Tunneling is found to be {\it approximately} momentum conserving in the sense that it has a Gaussian dependence on the momenta of the initial and final states. The magnitude of the coupling is determined by the intertube separation, and the crossing angle (which determines the contact area of the junction).

The developed model allows for a clear understanding of the effects of the different junction parameters on its low energy electronic properties. We find that the intrinsic asymmetries of the junction create a discrepancy between the forward and backward hopping between the tubes. This defines a zero-field Hall-like conductance for this four-terminal device. Using a Landauer-B{\"{u}}ttiker formalism, we calculate the different conductances of the four-terminal junction. We find that the contact conductance scales inversely with the zero-field Hall conductance of the junction.

We also find that, in general, the crossing-angle dependence of the junction contact conductance $G_c$ exhibits three peaks over the angular range. These peaks correspond to angles where matching between the tube lattices is highest, thereby maximizing the coupling. The relative magnitude of these peaks is understood in terms of the effective contact area between the tubes. Relative orientation of the tube lattices is found to affect the strength of the coupling between the tubes, though its effect on the contact conductance is small compared to that of the crossing angle. We also find that the contact conductance varies with the Fermi level of the junction, but also in a way which is less dramatic than that of the crossing angle. We therefore conclude that the two crucial parameters in determining the contact conductance are the tube chiralities and the crossing angle. 

We believe that this work has important implications on studies involving CNT networks, where it is customary to assume a fixed contact conductance between various tubes in a network. With our model, and given a certain distribution of chiralities in the network, one can study the effect of the chirality distribution on the CNT network resistance. This may prove that some chirality distributions are favored over others.

\section{acknowledgment} 
We are deeply indebted to C.L. Kane at the University of Pennsylvania for valuable discussions and suggestions. We thank M.A. Kuroda and G.J. Martyna at IBM Watson Research Lab for useful comments. This work is supported in part by Department of Energy under grant DE-FG02-ER45118 (EJM).

\bibliography{mybib}{}

\end{document}